
\input harvmac.tex
\rightline{FTUV/94-30}
\rightline{hep-th/9406114}
\rightline{June 1994}
\vskip 2truecm
\centerline{\bf QUANTUM HEISENBERG GROUP AND ALGEBRA:}
\vskip .5truecm
\centerline{\bf CONTRACTION, LEFT AND RIGHT REGULAR REPRESENTATIONS}
\vskip 2truecm
\centerline{{\bf Demosthenes  Ellinas}\foot{e-mail: ellinas@evalvx.ific.uv.es ;
supported by  DGICYT Spain }
and
{\bf Jan Sobczyk}\foot{permanent address: Institute
of Theoretical Physics, Wroc\l aw University, Pl. Maxa Borna 9, 50-205,
Wroc\l aw, Poland; e-mail: jsobczyk@plwruw11.bitnet ;
supported by EEC grant ERBCIPACT920488}
}
\vskip .5truecm
\centerline{Departamento de F\'isica Te\'orica and IFIC}
\centerline{Centro Mixto Universidad de Valencia, CSIC}
\centerline{ E-46100 Burjassot, Valencia, Spain}
\vfill
\noindent
\centerline{ABSTRACT}
\medskip
We show that the quantum Heisenberg group $H_{q}(1)$ can be obtained by
means of
contraction from quantum $SU_q(2)$ group.
Its dual Hopf algebra is the quantum Heisenberg algebra $U_{q}(h(1))$.
We derive left and right
regular representations for $U_{q}(h(1))$ as acting on its
dual $H_{q}(1)$. Imposing conditions on the right
representation the left representation is reduced to an irreducible
holomorphic representation with an associated quantum coherent state.
By duality,  left and right regular
representations for quantum Heisenberg group with the quantum
Heisenberg algebra as representation module are also constructed.
As before reduction of left representations
leads to finite dimensional
irreducible ones for which the intertwinning operator is also investigated.
\vfill
\eject

{\bf 1.} Following the advent of the so called q-deformed oscillator
\ref\mac{A. J. Macfarlane, J. Phys. {\bf A22 } 4581 (1989)},
\ref\bie{L. C. Biedenharn, J. Phys. {\bf A22} L873 (1989) },
\ref\sun{C-P. Sun and H-C. Fu, J. Phys. {\bf A22} L983 (1989) },
\ref\ck{M. Chaichian and P. Kulish, Phys. Lett. {\bf B234} 72 (1990) }
some alternatives were put forward for deformation
of the oscillator group
or
Heisenberg algebra
\ref\itaa{E. Celeghini, R. Giachetti, E. Sorace and M. Tarlini, J. Math. Phys.
{\bf 31} 2548 (1990)},
\ref\itab{
E. Celeghini, R. Giachetti, E. Sorace and M. Tarlini, J. Math. Phys.
{\bf 32} 1155 (1991)}.
The basic
idea of that scheme was a generalized version of the standard Wigner-In\"on\"u
contraction technique of Lie algebras, which starting from the quantum
$U_{q} (su (2))$ algebra could lead to one-dimensional Heisenberg algebra
$U_{q} (h (1))$. The corresponding contraction of the universal $R$-matrix
and its resulting fundamental matrix representation was further employed
to introduce the Heisenberg quantum matrix pseudogroup $H_{q} (1)$
which also was possessing a non trivial *-Hopf algebra structure.

One the other hand, recent developments in the representation theory of
quantum groups have provided explicit construction for the left and
right regular representations  of some classes of quantum algebras
e.g. $su_{q}(1,1)$
\ref\jap{T. Masuda, K. Mimachi, Y. Nakagami, M. Noumi, Y. Saburi
and K. Ueno, Lett. Math. Phys. {\bf 19} 187 (1990); ibid. 195 (1990)  },
quantum Lorentz algebra
\ref\ddf{
L. D\c{a}browski, V.K. Dobrev and R. Floreanini, J. Math. Phys.
{\bf 25} (1994) 971},
$sl_{q} (2)$ and $e_{q} (2)$
\ref\ds{
L. D\c{a}browski and J. Sobczyk, Lett. Math. Phys., in press.
},
$sl_{q} (3)$
\ref\dapa{
L. D\c{a}browski and P. Parashar, {\it Left Regular Representations of $Sl_q
(3)$:
Reduction and Intertwinners} , preprint, SISSA 57/94/FM.}
and
$sl_{q} (n)$
\ref\doba{
V. K. Dobrev, {\it q-difference Intertwining Operators for
$U_{q} (sl (n))$ : General Setting and the Case n = 3},
preprint ASI-TPA/10/93}.
Interesting contributions  by these
works include, the extension to the case of quantum algebras of the
notion of differential operators intertwining representations
\ref\dobb{
V.K. Dobrev, Rep. Math. Phys. {\bf 25} (1988) 159}
and some reduction schemes of the module of representation of the corresponding
algebras to holomorphic functions which could turn out to be useful in the
elucidation
of the relation between representation theory and geometry of
quantum groups.

In the present work we shall be concerned with
the quantum Heisenberg group and its algebra. This is an example of
fundamental importance
 both  for  potential applications
of quantum group theory and from its pedagogical character.
Here a new quantization procedure of the Heisenberg group
 $H_{q} (1)$, will be introduced by means of a contraction
scheme operating on the quantum group $SU_{q} (2)$, and its *-Hopf
algebra. Then the quantum Heisenberg algebra $U_{q} (h(1))$
is the dual Hopf algebra of  $H_{q} (1)$ .
Then we shall construct the
left and right regular representations of the $U_{q} (h (1))$ algebra
generators acting on the algebra of functions $H_{q} (1)$, taken as
the representation module. Next a reduction to an irreducible representation
submodule will be made, which will provide a holomorphic representation
for the $U_{q} (h (1))$ generators, to be regarded as quantum analogue of
the Bargmann representation. The eigenfunction of the left annihilation
 operator will also be obtained as a quantum
analogue of the canonical coherent states. Moreover, on duality
grounds, the  left and right regular representations of the quantum Heisenberg
group acting on its dual quantum Heisenberg algebra can also be  determined
by employing techniques analogous to those applied in the case of the
algebra representations.
In this case also reduction of the regular group representations to
finite dimensional ones and
their intertwinner operators will be found. Conclusions and some of the
perspectives
can be found at the end of the paper.

{\bf 2.} The quantum Heisenberg group $H_{q} (1)$ can be obtained by
means of
contraction method from the quantum $SU_{q} (2)$ group. The latter
\ref\frt{
L. D. Faddeev, N. Yu. Reshetikhin and L. A. Takhtajan, Leningrad Math. J.
{\bf 1} (1990) 193},
\ref\wor{
S. L. Woronowicz, Publ. R. I. M. S. {\bf 23} 117 (1987)},
\ref\vs{
L. L. Vaksman and Ya. S. Soibel'man, J. Fun. Anal. Appl. {\bf 22} 170 (1988)},
is defined by the relations ($T=\pmatrix{a & b\cr c & d}$)
\eqn\rel{
R T_{1} T_{2} = T_{2} T_{1} R
}
where
\eqn\rmat{
R = \pmatrix{
q & 0 & 0 & 0 \cr
0 & 1 & 0 & 0\cr
0 & q-q^{-1} & 1 & 0\cr
0 & 0 & 0 & q}
}
is the R-matrix, while the co-multiplication
$\Delta: SU_{q}(2) \rightarrow SU_{q}(2)^{\otimes^2}$
is defined by
\eqn\cop{
\Delta T_{in} = \sum_{k} T_{ik} \otimes T_{kn}
}
and the co-unit
\eqn\cou{
\varepsilon (T_{ij}) = \delta_{ij}
}
and antipode
\eqn\ant{
S T_{ij}    = T_{ij}^{-1}
}
together with the *-conjugation operation $T_{ij}^{*} = S T_{ji}
\quad , \quad q^{*} = q$, determine the *-Hopf algebra structure
of $SU_{q}(2)$ group, for which in addition the reality conditions,
\eqn\real{
d=a^{*} \ \ \ c=-q^{-1}b^{*}
}
and the determinant condition,
\eqn\det{
Det_{q}\equiv ad-qbc=aa^{*}+bb^{*}=1,}
are valid.

Let us assume that  element $d$ is invertible \ref\com{Using the determinant
and reality conditions we can write $d^{-1}=\sum_{0}^{\infty}(-)^{m}b^{m}
b^{*m} a$.
Careful consideration should be given
to this formula concerning its convergency, which should require some
completion  of the algebra.}.
Then $a$ can be expressed in
terms of $b, c, d$.
The substitutions $d = e^{l\beta }$,  $b = -l^{1/2} \alpha$,
$c = l^{1/2} \delta$ and $q = e^{\omega l}$ leads in the limit $l \rightarrow
0$, to the following commutation relations for elements $\alpha$, $\beta$
and $\delta$
\eqn\heal{
[\alpha, \delta ]  = 0,\ \ \ [\delta ,  \beta ] = \omega \delta ,\ \ \
[\alpha, \beta ] = \omega \alpha
}
and co-multiplication
$\Delta :H_{q}(1)\rightarrow H_{q}(1)^{\otimes^2}$,
\eqn\demlen{
\Delta (\beta)= \beta \otimes {\bf 1} + {\bf 1} \otimes \beta -
\delta \otimes \alpha
}
\eqn\hecop{
\Delta (\delta) = \delta \otimes {\bf 1} + {\bf 1} \otimes \delta\ \ \
\Delta (\alpha) = \alpha \otimes {\bf 1} + {\bf 1} \otimes \alpha
}
with antipode,
\eqn\hean{
S (\delta ) = - \delta ,\ \ \ S (\alpha) = - \alpha ,
\ \ \ S (\beta) = - \beta - \alpha \delta
}
and co-unit,
\eqn\heunit{
\varepsilon (\alpha) = \varepsilon (\beta) = \varepsilon (\delta) = 0 .
}
These are precisely the relations obtained in
\itab\
up to a  redefinition of the algebra generators, viz.
$ \omega \rightarrow {\omega}/2$, $\alpha \rightarrow - \delta$,
$\delta \rightarrow \alpha$.
Moreover the *-conjugation can also be obtained by means
of this contraction and it reads,
\eqn\hestar{
\delta^{*} = \alpha ,\quad \beta^{*} = - \beta - \alpha \delta .
}
which enjoys the property
\eqn\stardef{
(*\otimes *)\circ \Delta = \Delta\circ *
}
Altogether we obtain the *-Hopf algebra $H_{q} (1)$ which is dual to
the corresponding quantum Heisenberg algebra $U_{q} (h (1))$ \itab .
The latter is defined by the commutators,
\eqn\halal{
[H, A] = 0, \ \ \ [H, A^{+}] = 0, \ \ \ [A, A^{+}] = {sinh (\omega H)\over
\omega},
}
co-multiplication
$\Delta:H_{q}(1)\rightarrow H_{q}(1)^{\otimes^2}$,
$$\Delta(H) = H \otimes {\bf 1} + {\bf 1} \otimes H$$
\eqn\halcop{
\Delta (A) = A \otimes e^{\omega H/2} + e^{- \omega H/2} \otimes A, \ \ \
\Delta (A^{+}) = A^+ \otimes e^{\omega H/2} + e^{- \omega H/2} \otimes A^+
}
and antipode and co-unit given by
\eqn\halant{
S (X) = - X, \ \ and  \ \varepsilon (X) = 0, \quad X \in \{ A, A^{+}, H \} .
}
It possesses a R-matrix obtained by contraction of the $U_{q} (su (2))$ \itab .

{\bf 3.} Left and right actions of the quantum enveloping algebra
${\cal A}\equiv U_q(h(1))$
on its dual quantum group ${\cal A}^*\equiv H_q(1)$ are defined respectively by
$L:{\cal A}\times {\cal A}^*\rightarrow
{\cal A}^*$,  $(a,\phi )\rightarrow L(a)\phi$,

\eqn\lede{
(L(a)\phi ) (b)\equiv <b, L(a)\phi >=<S(a)b, \phi >\equiv \phi (S(a)b)
}
and $R: {\cal A}\times {\cal A}^*\rightarrow {\cal A}^*$,
$(a, \phi )\rightarrow R(a)\phi$,
\eqn\ride{
(R(a)\phi ) (b)\equiv <b, R(a)\phi >=<ba, \phi >\equiv \phi (ba)
}
where $a,b\in {\cal A}$ and $\phi, \psi \in {\cal A}^*$.
Using the properties of the duality pair (we use the notation
$\Delta \phi = \sum_{(\phi)} \phi_{(1)}\otimes \phi_{(2)}$)

\eqn\leca{
<b, L(a)\phi >=<S(a)b, \phi >= \sum_{(\phi)} <S(a), \phi_{(1)}> <b, \phi_{(2)}>
}
and
\eqn\rica{
<b, R(a)\phi >=<ba, \phi >= \sum_{(\phi)} <b, \phi_{(1)}> <a, \phi_{(2)}>,
}
the twisted derivation rule for the left action
\eqn\twde{
L(a)\phi\psi = \sum_{(a)} L(a_{(2)})\phi L(a_{(1)})\psi ,
}
the derivation rule for the right action
\eqn\twre{
R(a)\phi\psi = \sum_{(a)} R(a_{(1)})\phi R(a_{(2)})\psi ,}
and by virtue of the pairing,
$$< A^{n} A^{+l} H^{m}, \beta > = \delta_{k,0} \delta_{1,0} \delta_{m,1},$$
$$< A^{n} A^{+l} H^{m}, \delta > = \delta_{k,0} \delta_{1,1} \delta_{n,0},$$
\eqn\dudef{
< A^{n} A^{+l} H^{m}, \alpha > = \delta_{k , 1} \delta_{1,0} \delta_{m,0} .
}
and of the properties\ref\abe{
E. Abe, {\it Hopf Algebras}, Cambridge Tracts in Math. No. 74
(Cambridge U. P. 1980)
}
$${< X Y, \phi > = < X \otimes Y, \Delta( \phi) >},$$
\eqn\prop{
< X, \phi \psi > = < \Delta(X), \phi \otimes \psi > ,
}
$X, Y \in {\cal A}$ and ${\phi}, {\psi} \in {\cal A}^*$,
valid for dual pairs of Hopf algebras, we arrive at
following left and right regular action of the algebra:
$$L(A)\beta^n\alpha^m\delta^l=-m(\beta-{\omega\over 2})^n\alpha^{m-1}
\delta^l,$$
$$L(A^+)\beta^n\alpha^m\delta^l=-l(\beta-{\omega\over 2})^n\alpha^{m}
\delta^{l-1} + P_{(n)}(\beta )\alpha^{m+1}\delta^l,$$
\eqn\left{
L(H)\beta^n\alpha^m\delta^l=-n\beta^{n-1}\alpha^{m}
\delta^l
}
and
$$R(A^+)\beta^n\alpha^m\delta^l=l(\beta-{\omega\over 2})^n\alpha^{m}
\delta^{l-1},$$
$$R(A)\beta^n\alpha^m\delta^l=m(\beta-{\omega\over 2})^n\alpha^{m-1}
\delta^{l} - P_{(n)}(\beta )\alpha^{m}\delta^{l+1},$$
\eqn\right{
R(H)\beta^n\alpha^m\delta^l=n\beta^{n-1}\alpha^{m}
\delta^l .
}
These expressions could also have been obtained from $su_q(2)$ by
a contaction as in the case of $e_q(2)$ \ds .In the above formulae
\eqn\pol{
P_{(n)}(\beta )=\sum_{j=0}^{n-1}(\beta -{\omega\over 2})^{j}
(\beta +{3\omega\over 2})^{n-1-j} .
}
Due to the property
\eqn\pro{
{dP_{(n)}(\beta)\over d\beta} = n P_{(n-1)}(\beta ) ,
}
valid for $n\geq 2$ for $P_{(n)}$ taken as a function of $\beta$, it turns out
to be convenient to work in the following basis of $H_q(1)$: $x(r,m,l)
\equiv e^{r\beta}\alpha^m\delta^l$ where $r\in {\bf Z}, m,l\in {\bf N}$. We
observe that
\eqn\prob{
\sum_{n=1}^{\infty} {r^n\over n!}P_{(n)}(\beta ) = C_re^{r\beta } ,
}
where
\eqn\con{
C_r\equiv {1\over 2\omega} (e^{3\omega r\over 2} - e^{-{\omega r\over 2}}) .
}
In this new basis the left and right regular actions are taken respectively
the following forms,
$$L(A)x(r,m,l)=-me^{-{r\omega\over 2}}x(r,m-1,l),$$
$$L(A^+)x(r,m,l)=-le^{-{r\omega\over 2}}x(r,m,l-1) + C_rx(r,m+1,l),$$
\eqn\lere{
L(H)x(r,m,l)=-rx(r,m,l)
}
and
$$R(A^+)x(r,m,l)=le^{-{r\omega\over 2}}x(r,m,l-1),$$
$$R(A)x(r,m,l)=me^{-{r\omega\over 2}}x(r,m-1,l) - C_rx(r,m,l+1),$$
\eqn\rire{
R(H)x(r,m,l)=rx(r,m,l) .
}
One can easily verify that $L$ and $R$ are in fact representations
of the $U_q(h(1))$ and that $[R(X), L(Y)]=0$ for $X,Y\in \{A,A^+,H\}$.

We turn now to the reduction of the above representations. First we observe
that $L(A)$ and $R(A^+)$ act like "annihilation" operators and their action is
bounded from below. The "vacuum" condition $R(A^+)=0$ implies $l=0$
and analogously $L(A)=0$
implies $m=0$. As we want to obtain an irreducible left representation
(right representation reduction goes along same lines),
we choose to
impose on the representation space the condition $l=0$ as well as the condition
that
the action of $R(H)$ is just a multiplication by a chosen integer, say $s$.
This condition fixes $r$ to be $r=s$. Under this conditions the form of the
left regular representation becomes
$$L(A)x(s,m,0)=-me^{-{s\omega\over 2}}x(s,m-1,0),$$
$$L(A^+)x(s,m,0)=C_sx(s,m+1,0),$$
\eqn\rele{
L(H)x(s,m,0)=-sx(s,m,0) .
}

We see that the label $s$ (the term $e^{s\beta}$) remains unchanged under
the left action of our algebra.  It is therefore natural to pass to the
induced representation on monomials (and then by linearity to holomorphic
functions)
$\alpha^m$,  $m\in {\bf N}$. Allowing the formal derivation with respect to
$\alpha$ (strictly speaking it is an operator) we arrive at a holomorphic
realization of the quantum Heisenberg algebra generators:

$$L(A)=-e^{-{s\omega\over 2}}{\partial \over \partial\alpha},$$
$$L(A^+)=C_s\alpha,$$
\eqn\hol{
L(H)=-s .
}
If we introduce the following elements of the representation space
$|s; m>\equiv {\alpha^m\over \sqrt{m!}}$ we get
$$L(A) |s; m>= -\sqrt{m}e^{-{s\omega\over 2}} |s; m-1>,$$
$$L(A^+) |s; m>= \sqrt{m+1} C_s |s; m+1>,$$
\eqn\bra{
L(H) |s; m>= -s |s; m> .
}
Finally we can redefine elements $A, A^+, H$ in the following way
(here we have to assume that $C_s\neq 0$)
$$\tilde H = - H,$$
$$\tilde A = - e^{s\omega\over 2}\sqrt{sinh(\omega s)\over \omega} A,$$
\eqn\redef{
\tilde A^+ = {1\over C_s}\sqrt{sinh(\omega s)\over \omega} A^+  .
}
This transformation does not change the quantum Heisenberg algebra
relations \halal . Moreover we get the following expressions for the
left representation

$$L(\tilde A )|s; m> = \sqrt{m{sinh(\omega s)\over \omega}}|s; m-1>,$$
$$L(\tilde A^+ )|s; m> = \sqrt{(m+1){sinh(\omega s)\over \omega}}|s; m+1>,$$
\eqn\ost{
L(\tilde H )|s; m> = s |s; m>,
}
which are exactly the same as those obtained in \itaa . Let us also mention for
completeness that the operator which in the case of semisimple algebras
plays a role of interwinner takes here the form
\eqn\era{
R(A)x(r,m,l)= me^{-{r\omega\over 2}}x(r,m-1,l) - C_rx(r,m,l+1) .
}
This introduces an unwanted operator $\delta$, thus leading us outside the
subspace of elements defined by the condition $l=0$.
Finally we observe that the analytic vectors defined as (${\mu}^2 \equiv
{sinh(s\omega )\over \omega }$),
\eqn\anve{
|\alpha >\equiv
e^{\alpha^2\over 2\mu} ,
}
for each irreducible left regular representation labelled by $s$,
determine eigenfunctions of the annihilation operator i.e.
\eqn\coh{
L(\tilde A)|\alpha > = \alpha |\alpha >
}
and define a formal deformed canonical unnormalized coherent state for
the
quantum Heisenberg algebra
\ref\chek{
M.Chaichian, D.Ellinas and P.Kulish, Phys. Rev. Lett. {\bf 65} (1990) 95},
\ref\ella{D.Ellinas, J. Phys. {\bf A26} (1993) L51}.

{\bf 4.} As mentioned in the introduction  proceeding in analogous way as
in the case of the quantum algebra we can find a representation of the quantum
Heisenberg group acting on its dual quantum Heisenberg algebra. We obtain the
following expressions for the left regular representation of $H_{q}(1)$ ,
$$L(\alpha )H^k(A^+)^mA^n = -n H^k e^{H\omega \over 2} (A^+)^mA^{n-1},$$
$$L(\delta )H^k(A^+)^mA^n = -m H^k e^{H\omega \over 2} (A^+)^{m-1}A^{n},$$
\eqn\grle{
L(\beta )H^k(A^+)^mA^n = -k H^{k-1} (A^+)^mA^{n}
+{\omega (m+n)\over 2} H^k(A^+)^mA^n ,
}
while its right regular representation reads,
$$R(\alpha )H^k(A^+)^mA^n = n H^k e^{-{H\omega \over 2}} (A^+)^mA^{n-1},$$
$$R(\delta )H^k(A^+)^mA^n = m H^k e^{-{H\omega \over 2}} (A^+)^{m-1}A^{n},$$
\eqn\grri{
R(\beta )H^k(A^+)^mA^n = k H^{k-1} (A^+)^mA^{n}
+{\omega (m+n)\over 2} H^k(A^+)^mA^n - mnH^ke^{H\omega}(A^+)^{m-1}A^{n-1} .
}
Of course in the limit as $\omega\rightarrow 0$ $\alpha , \beta , \delta $
become commuting among themselves.
Applying the procedure of reducing the $L$ representation by imposing some
conditions on $R$, we can arrive at finite dimensional irreducible
representations of the Heisenberg group, which we recall that
as an algebra it is a Lie
algebra but has non trivial bialgebra structure.
 To this end we first introduce a more convenient
basis in the quantum Heisenberg algebra consisting of elements
$y(k,m,n)\equiv e^{k\omega H\over 2}(A^+)^mA^n$. In this basis the
left and right action become correspondingly,
$$L(\alpha )y(k,m,n) = -n y(k+1,m,n-1),$$
$$L(\delta )y(k,m,n) = -m y(k+1,m-1,n),$$
\eqn\grlen{
L(\beta )y(k,m,n) = {\omega\over 2}(m+n-k) y(k,m,n)
}
and
$$R(\alpha )y(k,m,n) = n y(k-1,m,n-1),$$
$$R(\delta )y(k,m,n) = m y(k-1,m-1,n),$$
\eqn\grrin{
R(\beta )y(k,m,n) = {\omega\over 2}(m+n+k) y(k,m,n) -mn y(k-2,m-1,n-1) .
}
We first impose the condition $R(\alpha )=0$, which fixes $n=0$. Then we
demand that all the vectors of the representation space are $R(\beta )$
eigenstates to the eigenvalue ${p\omega\over 2}$ for some integer $p$.
This imposes a constraint
on possible values of $k$ and $m$ as $k+m=p$. The action of elements of left
representation and of $R(\delta )$ becomes (we use a notation
$|p; m>\equiv y(p-m,m,0)$)
$$L(\alpha ) |p; m> = 0,$$
$$L(\delta ) |p; m> = -m |p; m-1>,$$
\eqn\grlef{
L(\beta ) |p; m>  = {\omega\over 2}(2m-p) |p; m>
}
and
\eqn\grif{
R(\delta ) |p; m>  = |p-2; m-1> .
}
If we do not impose any restrictions on $m$ it is clear that \grlef\
are infinite dimensional irreducible representations of the quantum
Heisenberg group and representations are labelled by an integer $p$. Then
$R(\delta )$ can be viewed as an intertwinner as it is a map between
representations labelled by $p$ and $p-2$. We can write it as
\eqn\inter{
L_{p-2}(X)R_{p}(\delta ) = R_{p}(\delta )L_{p}(X), }
where we have added subscripts $p$ and $p-2$ to operators $L$ and $R$ ($X$ can
be
any generator of the Heisenberg group) in order to make clear on which
representation space do they act. Finally we observe that for given $p$
the set of
vectors in $ker(R(\delta )^M)$ form an invariant subspace
under the action of $L$'s and thus provide a $M$-dimensional irreducible
representation of the quantum Heisenberg algebra.

{\bf 5.} In conclusion, we have presented a construction of
the Hopf
algebra structure of the quantum Heisenberg group by a
the contraction method; this group is dual as a Hopf algebra to the quantum
Heisenberg algebra. Also regular representations of the quantum Heisenberg
group and algebra where obtained together with intertwinners.
Reduction schemes of regular representations to holomorphic
realizations for quantum Heisenberg algebra where also provided and
the associated quantum analogues of canonical coherent states where
pointed out. There are a number of interesting topics to pursue
further: given the R-matrix of the quantum Heisenberg group,
a quantum differential calculus can be developed and studied;
additionally employing the contraction method, the algebraic integral calculus
available for $SU_{q}(2)$ could be induced into the Heisenberg group and
used further to establish hermiticity of the regular representations
and to develop a notion of square integrability in the modules of these
representations; to these and related problems we aim to return elsewhere.

{\bf Acknowledgement:} We should like to thank Prof. J. A. de Azc\'arraga for
discussions. One of the authors (JS) thanks also for a warm hospitality
at Departamento de Fisica Teorica, Universidad de Valencia, where this
work was done.
\listrefs
\end